\documentclass[12pt]{article}
\usepackage{amsfonts}
\usepackage{graphicx,amssymb,bm,makeidx}
\usepackage[small,bf]{caption2}

\newcommand{\I}{{\rm i}}

\renewcommand{\Im}{{\rm Im}}

\begin{document}

\title{Magnetic fluctuations and itinerant ferromagnetism in two-dimensional systems with van
Hove singularities}
\author{P. A. Igoshev$\mathrm{^{a)}}$, A. A. Katanin$\mathrm{^{a),b)}}$, and
V. Yu. Irkhin$\mathrm{^{a)}}$}
\maketitle

\noindent $\mathrm{^{a)}}$ Institute of Metal Physics 620041, Ekaterinburg,
Russia \newline
$\mathrm{^{b)}}$ Max-Planck-Institut f\"ur Festk\"orperforschung, 70569
Stuttgart, Germany


\begin{center}
{\large Abstract }
\end{center}

We use the quasistatic approach to analyze the criterion of ferromagnetism for two-dimensional (2D) systems with the Fermi level near Van
Hove singularities (VHS) of the electronic spectrum. It is shown that the spectrum of spin excitations (paramagnons) is positively
defined when the interaction $I$ between electrons and paramagnons, which corresponds to the  Hubbard on-site repulsion $U$, is sufficiently large. The critical
interactions $I_c$ and $U_c$ remain finite at Van Hove filling and exceed considerably their values obtained from the Stoner
criterion due to incommensurate spin fluctuations, which are
important near the ferromagnetic quantum phase transition. Combining
the quasistatic approximation and the equation of motion method for the
Green function we obtain the results for the electronic self-energy to
first order in the inverse number of spin components.


\newpage

\section{Introduction}

{
Two-dimensional and layered systems with strong electronic correlations have
been attracting attention of researchers for more than two decades. In such
systems, in particular, the $\mathrm{CuO_2}$ planes of the $\mathrm{La_{2-\mathit{x}}Sr_{%
\mathit{x}}CuO_4}$ and $\mathrm{YBa_2Cu_3O_{7-\delta}}$ compounds the low
dimensionality favors appearance of the high-$T_c$ singlet superconductivity
\cite{DCV}. On the other hand, a number of low-dimensional triplet type
superconducting compounds (Sr$_2$RuO$_4$, UGe$_2$) have been discovered
recently. For example, the compound UGe$_2$ is ferromagnetic at low temperatures and
pressure. It is expected that the ferromagnetic fluctuations may play a
crucial role in the properties of the paramagnetic compound Sr$_2$RuO$_4$
\cite{Mazin}. This is confirmed by the properties of the
La-doped compound La$_x$Sr$_{2-x}$RuO$_4$ which is near the
ferromagnetic transition at $x=0.27$ \cite{ED}, and also by the properties of
the isoelectronic compound Ca$_2$RuO$_4$ \cite{Maeno1}, which becomes
ferromagnetic under pressure. The above-mentioned compounds have Van Hove
singularity (VHS) of the density of states near the Fermi level,
the description of magnetic fluctuations in the presence of these
singularities is an important problem, which may serve as a basis for deeper
insight into magnetic and superconducting properties of these compounds.}

According to the Stoner theory, the large value of the density of states at
the Fermi level, occurring due to VHS, leads to a possibility of
ferromagnetically ordered ground state. The Stoner theory, however, neglects
fluctuations of magnetic order parameter, which may significantly change the
necessary conditions for ferromagnetism.
It is well known that for systems with nonsingular density
of states the Stoner theory does not explain correctly magnetic and thermodynamic properties, e.g. it predicts larger temperatures of magnetic
phase transition in comparison with the experimental data.
Starting from the papers by Murata, Doniach \cite{Murata} and Dzyaloshinskii
\cite{Kondr}, the paramagnon theory has been developed to account for the effect of spin fluctuations and describe the properties of weak- and nearly ferromagnetic materials. Later on this theory
was systematically formulated by Moriya \cite{Moriya}.  This theory considers magnetic excitations (paramagnons), which give the dominant contribution to
thermodynamic properties of weak ferro- and antiferromagnets. Properties of paramagnons in the
vicinity of quantum phase transitions were investigated later within the
renormalization group method by
Hertz in the framework of the $\phi ^4$-model \cite{Hertz}, the corresponding results were generalized by Millis to
finite temperatures \cite{Millis}. The applicability of the
Hertz-Moriya-Millis theory in two- and three dimensions has been, however, recently questioned, because the lowest order corrections to the spin susceptibility
were shown to be non-analytical with respect to momentum and therefore dramatically change the
spectrum of magnetic excitations \cite{Chubukov}.


In the presence of VHS in the vicinity of the Fermi level the Hertz-Moriya-Millis
theory itself, however, is not applicable to describe the ferromagnetic ground state. 
Indeed,
due to the influence of VHS the noninteracting magnetic susceptibility $\chi_q^0$ may not
have a maximum at the point ${q=0}$ \cite{Onufr}. Since the spectrum of
paramagnons in the Hertz-Moriya-Millis theory is determined by the inverse
spin susceptibility $(\chi_q^0)^{-1}$, this spectrum is not positively defined in the presence of VHS and the ground state with the spin density wave at the wavevector $\mathbf{Q}\ne0$
becomes more preferrable in this theory. Due to the presence of the electron-electron interaction, however, the paramagnon subsystem is not independent, but it is interacting with the electronic degrres of freedom, which can change the position of the spin susceptibility maximum in momentum space. Since the Hertz-Moriya-Millis theory neglects the
renormalization of the momentum dependence of the susceptibility
by the electron-paramagnon interaction, it cannot therefore describe correctly ground state properties in the presence of VHS.
Furthermore, for the position of the Fermi level in the vicinity of the VHS
the paramagnon interaction energy in the Hertz-Moriya-Millis theory is
\textit{negative}, which
corresponds to attraction between paramagnons with the possibility of bound
states formation.  The magnitude of this interaction may be also changed
essentially by higher-order terms in the electron-paramagnon interaction $I$, which can result in the repulsive paramagnon interaction.

Therefore, to investigate the possibility of the ferromagnetic ordering in the
presence of VHS it is necessary to take into account the effect of the
electron-paramagnon interaction on the spin susceptibility more accurately than in the
existing spin-fluctuation theories. An attempt to go beyond these theories
was made by Hertz and Klenin \cite{Klenin}, who considered the sum of
infinite series of diagrams for irreducible susceptibility.
However, the renormalization of
the momentum dependence of the susceptibility, which becomes crucial in the
presence of VHS, was not considered in their study. The aim of the present paper is to formulate a consistent spin-fluctuation theory in the presence of VHS and to investigate their influence on the possibility of ferromagnetism.

To this end, we consider the spin-fermion model, which treats interacting electronic and spin degrees of freedom. This model was introduced in early 1990's to treat antiferromagnetic
fluctuations in high-$T_c$ superconductors \cite{SF} (see also Refs. \cite{Schmalian,QS}), but can be generalized
to the case of ferromagnetic fluctuations \cite{Katanin}.
Since previous attempts to derive this model from the Hubbard model met the problem of the so-called Fritz ambiguity, we use alternative way to derive this model and obtain the relation between the parameters of the Hubbard and spin-fermion models. Having this relation, the phase diagram of these models can be investigated.  It is argued that for large enough electron-paramagnon interaction $I_c$
(and the corresponding Hubbard on-site Coulomb repulsion $U_c$)
the spectrum of paramagnons becomes positively defined, which corresponds to the
possibility of ferromagnetism at $I>I_c$. The quantities $I_c$ ($U_c$)
exceed substantially the corresponding values, obtained from the Stoner criterion. The
latter is in an agreement with the results of the functional group approach \cite
{Kampf_RG} and the numerical investigations for the Hubbard model. The critical values of the interaction $I_c (U_c)$ are finite at the Van Hove filling.

The plan of the paper is the following. In Section 2 we introduce the spin-fermion model and derive the ``generalized'' Stoner criterion for ferromagnetism. 
In Sections 3.1 and 3.2 the application of this criterion for  investigation of the phase diagram and magnetic properties of the model is presented. The  additional criterion of ferromagnetism, requiring the positivity of the energy of the paramagnon interaction is considered in Section 3.3 and showed to be less important than the generalized Stoner criterion. In Section 3.4 we combine the quasistatic approach with the equation of motion method for the electronic Green function which allows us to investigate electronic properties of the system. The obtained results are discussed in  Section 4. In Appendix A we discuss the derivation of the spin-fermion model from the Hubbard model. The diagram technique for the spin-fermion model is presented in Appendix B. The formulae for the second derivative of  magnetic susceptibility with respect to the momenta are presented in Appendix C.
\section{Magnetic Properties of Two-Dimensional Systems}

\subsection{The Hubbard model and an applicability of the Stoner criterion of ferromagnetism}



To investigate the effect of VHS on magnetic properties of 2D systems we
consider the Hubbard model
\begin{equation}
\mathcal{H}=\sum\limits_{\mathbf{k\sigma }}\varepsilon _{\mathbf{k\sigma }%
}c_{\mathbf{k}\sigma }^{+}c_{\mathbf{k}\sigma }+U\sum\limits_{i}n_{i\uparrow
}n_{i\downarrow }.  \label{Hubb}
\end{equation}
with the electronic dispersion
\begin{equation}
\varepsilon_{\mathbf{k}}=-2t(\cos k_x+\cos k_y)+4t^{\prime}(\cos k_x\cos
k_y+1)-\mu.  \label{ek}
\end{equation}
where $t$ and $t^{\prime}$ are the nearest- and next-nearest neighbor hopping
parameters, $U$ is the Hubbard on-site repulsion, $n_{i\sigma}=c^+_{i\sigma}c_{i\sigma}$.

\begin{figure}[tbp]
\noindent\includegraphics[width=0.5\textwidth]{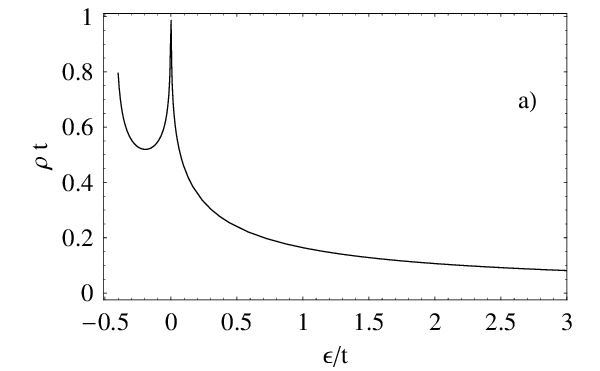} \includegraphics[width=0.5\textwidth]{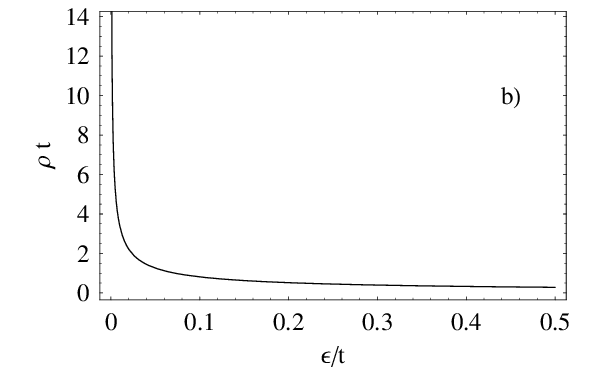}
\caption{\textsf{{\protect\footnotesize {Electronic noninteracting density of states $%
\protect\rho(\protect\varepsilon)$ at a) $t^{\prime}=0.45t$,
b) $t^{\prime}=0.50t$}}}}
\label{DOS}
\end{figure}

Let us first consider the noninteracting case. The corresponding plots of the non-interacting density of states for different $%
t^{\prime}$ are shown in Fig. \ref{DOS}. For $t^{\prime}<0.5t$ the density
of states has logarithmic singularity at the energy $4t-8t^{\prime}$
measured from the bottom of the band. For $t^{\prime}=0.5t$ there are lines
of VHS along the $k_x=0$ and $k_y=0$ directions, and the density of states
has stronger divergence $\rho(\varepsilon)\propto 1/\varepsilon^{1/2}$ at
the bottom of the band (the so-called flat band case \cite{Kampf_RG,Tasaki}),
which is analogous to the giant VHS in 3D systems \cite{Katsnelson}%
. The dynamical susceptibility of the noninteracting electronic gas with the
dispersion (\ref{ek}) is
\begin{equation}
\chi_q^{0}=- \frac{T}{N}\sum_k \mathcal{G}_k \mathcal{G}_{k+q}
\label{Pauli}
\end{equation}
where $\mathcal{G}_k=(\mathrm{i}\nu_n-\varepsilon_\mathbf{k})^{-1}$ is the
bare electronic Green function, ${k=(\mathbf{k};\mathrm{i}\nu_n)}$ and $q=(%
\mathbf{q};\mathrm{i}\omega_n)$, $\nu_n=(2n+1)\pi T$ and $\omega_n=2n\pi T$
are the Fermi and Bose Matsubara frequencies. In the presence of VHS the
uniform susceptibility is logarithmically divergent near Van-Hove band fillings with $\mu=0,$
\begin{equation}
\chi_0^{0}\simeq\frac1{2\pi t}\ln\frac{t}{\max(|\mu|,T)}.
\end{equation}
According to the Stoner criterion, this leads to a possibility of ferromagnetism for $U>U_c^{\rm Stoner}=1/\chi_0^0$.
However, the momentum
dependence $\chi_{(\mathbf{q};0)}^{0}$ at low $T\ll t$ and $\mu\neq 0$ has its maximum at $\mathbf{q}\ne0$
(see Ref. \cite{Onufr} and Figs. \ref{q-chi},\ref{q-chi2} with $\Delta=0$ below). 
In the random phase approximation (RPA)
\begin{equation}
\chi_q^{\rm RPA}=\frac{\chi_q^0}{1-U\chi_q^0}.
\end{equation}
this results in an instability of ferromagnetic ground state, since $\chi_{(\mathbf{q};0)}^{\rm RPA}$ is divergent at $\mathbf{q}\neq0$ even for $U<U_c^{\rm Stoner}$, which implies the spin-density wave instability in this approach. This shortcoming holds also
in the Hertz-Moriya-Millis theory, not considering the renormalization of momentum dependence of the electron-electron interaction and implies the impossibility of ferromagnetism near van Hove fillings in existing spin-fluctuation approaches.
\subsection{The spin-fermion model}

For the purpose of investigation of magnetic properties beyond RPA-type approaches
we consider the \textit{%
spin-fermion model} \cite{SF}
\[
\mathcal{Z}_{\mathrm{sf}}[\eta ,\eta ^{+}]=\int D[\mathbf{S};c,c^{+}]\exp
(-\beta \mathcal{S}),
\]%
\begin{equation}
\mathcal{S}=\sum_{k\sigma }c_{k\sigma }^{+}(-\mathrm{i}\nu _{n}+\varepsilon
_{\mathbf{k}})c_{k\sigma }+\mathcal{S}_{S}+\frac{I}{N}\sum_{q}\mathbf{s}_{q}%
\mathbf{S}_{-q}-\sum_{k\sigma }\left( \eta _{k\sigma }^{+}c_{k\sigma
}+c_{k\sigma }^{+}\eta _{k\sigma }\right) ,  \label{Sc}
\end{equation}%
\[
\mathcal{S}_{S}=\frac{1}{N}\sum_{q}{\chi _{q}^{-1}}\mathbf{S}_{q}\mathbf{S}%
_{-q}.
\]%
where the Grassman fields $c_{k\sigma },c_{k\sigma }^{+}$ correspond to electronic degrees
of freedom, the field $%
\mathbf{S}_{q}$ describes the collective spin degrees of freedom,
corresponding to paramagnons, $\chi_{q}$ is the
``bare'' susceptibility of the paramagnon subsystem, $\mathbf{s}_{q}=\sum_{k\sigma \sigma ^{\prime }}\bar{c}_{k\sigma
}^{+}\vec{\sigma}_{\sigma \sigma ^{\prime }}c_{k+q\sigma ^{\prime }}$ is
the spin operator of itinerant electrons, $\vec{\sigma}$ are the Pauli matrices, $%
I $ is the interaction of the paramagnon and electron subsystems, $\eta _{k\sigma },\eta _{k\sigma }^{+}$ are the fermionic sources
fields,  $\beta =1/T$ is the inverse
temperature, and $N$ is the number of sites.

Although
the model (\ref{Sc}) was previously derived from the Hubbard model via the Hubbard-Stratonovich
transformation, this way of derivation meets the problem of the so-called Fritz ambiguity 
\cite{Schulz,Weng,LargeN,Bickers,Dupuis,Wetterich} and the resulting action is not uniquely defined.
 In this paper we use a different way to establish connection between the spin-fermion
and Hubbard model, and show that the former describes magnetic
degrees of freedom of the latter, see Appendix A. The equivalence with the Hubbard model requires
\begin{equation}
I=U, \ \chi _{q}^{-1}=(\chi^{\rm el}_q)^{-1}+U^{2}H_{q},\, 
\end{equation}
where $\chi^{\rm el}_q$ is the exact magnetic susceptibility of the Hubbard model, which is calculated self-consistently in the framework of spin-fermion model according to $(\chi^{\rm el}_q)^{-1}=H_q^{-1}-U$, $H_q$ is the  irreducible electronic polarization operator. In the most part of the paper we consider, however, the spin-fermion model with arbitrary $\chi _{q}$ and $I$.

To investigate the magnetic properties of the model (\ref{Sc}) it is convenient
to rewrite the action in terms of the bosonic fields only \cite{Klenin}.
Integrating out electronic
degrees of freedom and expanding the resulting functional in the series in $I$ we obtain 
\[
\mathcal{Z}_{\mathrm{sf}}[\eta ,\eta ^{+}]=\int D[\mathbf{S}]\exp \left[
-\beta \mathcal{S}_{S}-\beta \mathcal{S}_{\mathrm{int}%
}-\sum_{kk^{\prime }\sigma \sigma ^{\prime }}\eta _{k\sigma }^{+}\mathbb{G}%
_{k\sigma ,k^{\prime }\sigma ^{\prime }}\eta _{k^{\prime }\sigma ^{\prime }}%
\right] ,
\]%
\begin{equation}
\mathcal{S}_{\mathrm{int}%
}=T\sum_{n=2}^{\infty }\frac1{n}{\mathrm{Tr}\left[ \left( \frac{I}{N}\mathcal{G}_{k}%
\vec{\sigma}\mathbf{S}_{k-k^{\prime }}\right)^{n}\right]_{k\sigma,k'\sigma'} },  \label{SR}
\end{equation}%
where
\begin{equation}
\mathbb{G}_{k\sigma ,k^{\prime }\sigma ^{\prime }}\left[ \mathbf{S}\right]
=\left\{ (\mathrm{i}\omega _{n}-\varepsilon _{\mathbf{k}})\delta
_{kk^{\prime }}\delta _{\sigma \sigma ^{\prime }}-(I/N)\vec{\sigma}_{\sigma
\sigma ^{\prime }}\mathbf{S}_{k-k^{\prime }}\right\} _{k\sigma ,k^{\prime
}\sigma ^{\prime }}^{-1}  \label{G(S)}
\end{equation}%
is the Green function of electrons, propagating in the presence of an
external field $\mathbf{S}$, 
$\mathcal{S}_{\mathrm{int}}$ corresponds to paramagnon interaction, the exponent $n$ corresponds the power of matrix in the square brackets taken with respect to indices $k\sigma,k'\sigma'.$ 
Coefficients of the expansion of the interaction $\mathcal{S}_{%
\mathrm{int}}$ in powers of the spin fields $S_{k-k^{\prime}}^j$ ($j=x,y,z$)
determine the vertices of paramagnon interaction
\begin{equation}
\Gamma^{j_1\ldots j_r}_0(q_1,\ldots,q_r)=\frac{T I^r}{r} \sum_{\mathcal{P}%
_q}\sum_{k} \mathrm{Tr}_\sigma\left[ \prod_{i=1}^r (\mathcal{G}_{k+\sum
_{l=1}^i q_l} \sigma^{j_i})\right]  \label{Loop}
\end{equation}
($\mathcal{P}_q$ corresponds to all possible permutations of momenta $q_i$, $\mathrm{Tr}%
_\sigma$ is the trace with respect to spin variables).
Taking into account the fourth order vertex only leads to an effective $%
\phi^4$ model \cite{Hertz,Millis} with the paramagnon dispersion $(\chi_q^{\rm RPA})^{-1}$ and the interaction $\Gamma_0(q_1,q_2,q_3,q_4)$. Below we consider, however, the effect of infinite sequence of vertices (\ref{Loop}) on the magnetic properties.

Diagram technique for the model (\ref{SR}) is considered in Appendix B and
contains the following elements:

\begin{itemize}
\item the propagators of the longitudinal and transverse paramagnons $%
\mathcal{R}_q$;

\item the electronic Green functions $\mathcal{G}_k$, connected in loops (\ref{Loop});

\item the electron-paramagnon interaction vertices $I \vec{\sigma}$.
\end{itemize}
These rules are used below to calculate physically observable quantities.

\subsection{Spin susceptibility}

To investigate magnetic properties of the model (\ref{Sc}) we consider magnetic susceptibility, defined as the causal double-time Green function
\begin{equation}
\chi^{%
\mathrm{el}}_q=\frac1{4N}\int_0^{\beta}\, d\tau e^{\I \omega_n \tau} \langle T_\tau s_{\bf q}^{+}(\tau)s_{-{\bf q}}^{-}(0)\rangle
\end{equation}
and the paramagnon propagator
\begin{equation}%
\mathcal{R}_q=\frac1{N} \int_0^{\beta}\, d\tau e^{\I \omega_n \tau} \langle T_\tau S_{\bf q}^{+}(\tau)S_{-{\bf q}}^{-}(0)\rangle.
\end{equation}
Here $s_{q}^{\pm}(\tau)$ and $S_{q}^{\pm}(\tau)$ are the operators in the Heisenberg representation, corresponding to the $s_{q}^{\pm}$ and $S_{q}^{\pm}$ fields, $T_\tau$ denotes the chronological product with respect to the imaginary time. To obtain the result for
the susceptibility $\chi^{\mathrm{el}}_q$, one can differentiate
Eq. (\ref{SR}) with respect to $\eta$.
Classifying corresponding contributions to the one particle reducible and
irreducible, we obtain (see Appendix A)
\begin{equation}
\chi ^{\mathrm{el}}_{q}=\frac{H_q}{1-I^2H_q\chi_q}+\tilde{\chi} ^{\mathrm{el}}_{q},  \label{hiel3}
\end{equation}
where $\tilde {\chi} ^{\mathrm{el}}_{q}$ is the non-paramagnon contribution to the electronic susceptibility ($\tilde {\chi} ^{\mathrm{el}}_{q}=U H_q^2$ for the Hubbard model).
The result (\ref{hiel3}) reduces for $H_q=\chi_q^{0}$ to the RPA result for the spin susceptibility.

The general expression for the paramagnon propagator is
\begin{equation}
\mathcal{R}_q^{-1}=\chi_q^{-1}-\Xi_q, \label{chiHq}
\end{equation}
where $\Xi_q$ is the paramagnon self-energy.  
It may be established diagrammatically that this self-energy can be expressed through the polarization operator $H_q$ according to $\Xi_q=I^2H_q$ (this relation is in fact a consequence
of the form of the interaction term of the model (\ref{Sc})).

\begin{figure}[tbp]
\noindent\center\includegraphics[width=0.8\textwidth]{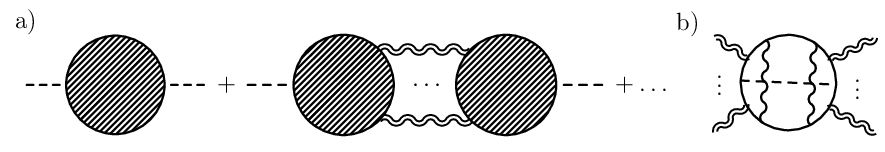}
\caption{\textsf{{\protect\footnotesize {(a) Series of diagrams for the
polarization operator of electronic subsystem; the dashed circle
represents the sum of all one-loop diagrams, as in Fig. 2b. (b) The example of the one-loop diagram, double wavy
lines correspond to the longitudinal or the transverse propagator,
denoted by dashed and wavy lines, respectively}}}}
\label{reducible}
\end{figure}

The polarization operator $H_q$ can be represented as a set of
diagrams, which contain loops of electronic Green functions
(the one-loop vertex functions, defined in Appendix B), connected by two or
more paramagnon lines (see Fig. 2). In the present paper we neglect
contributions to $H_q$, which contain more than one
electronic loop (second and next terms in Fig. \ref{reducible}a),
since we expect that their contribution to the magnetic susceptibility is small. The contribution of the diagrams with only
one electronic loop (Fig. 2b) corresponds the one-loop 2-point vertex
\begin{eqnarray}
H_q= \left\langle \Pi_q[\mathbf{S}%
]\right\rangle_0=:-\frac1{2I^2}\Gamma_{\mathrm{1-loop}}^{zz}(q,-q),
\label{Pi} \\
\Pi_q[\mathbf{S}]=-\frac{T}{2N}\sum_{k_1k_2} \mathrm{Tr}_\sigma \left[%
\mathbb{G}_{k_1,k_2}[\mathbf{S}]\sigma^z\mathbb{G}_{k_2+q,k_1+q}[\mathbf{S}%
]\sigma^z\right].  \nonumber
\end{eqnarray}

According to the Mermin-Wagner theorem \cite{Mermin}, the long range order
in 2D systems is possible at $T=0$ only.
The susceptibility  (\ref{hiel3})  in the limit $T\rightarrow
0$ corresponds to the susceptibilities of the ordered state, averaged over
directions:
\begin{equation}
\lim_{T\rightarrow 0}\chi^{\mathrm{{el}}}_q=\frac{1}{3}%
(\chi^{zz}_q+\chi^{+-}_q)^{\mathrm{{el}}}_{T=0}.
\end{equation}
The necessary condition of the ferromagnetic ground state is
\begin{equation}
I^2 \chi_0 H_{q=0}=1\; \mbox{at}\; T\rightarrow 0
\label{Stoner}
\end{equation}
(generalized Stoner criterion), and the positivity of the spectrum of
excitations of the static paramagnons $\omega_\mathbf{q}\equiv\mathcal{R}_{(\mathbf{q};0)}^{-1}$ at $T\rightarrow0$%
, which is fulfilled if the product $\chi_{(\mathbf{q};0)}H_{(%
\mathbf{q},0)}$ is maximal at $\mathbf{q}=0$. This criterion is violated in RPA, where $\omega_%
\mathbf{q}$ is not positively defined
in the $T\rightarrow0$ limit in the lowest (second) order of perturbation
theory in $I$ (see Section 2.1). Moreover, the corresponding energy
of the paramagnon interaction $\Gamma_0(0,0,0,0)=-12I^4\rho^{\prime\prime}(\mu)$
is also negative near VHS. To investigate the possibility of ferromagnetism we consider below the results for the polarization operator $H_q$ beyond RPA.




\subsection{The static and the quantum contributions. The static
approximation}


In the assumption of ferromagnetically ordered ground state considered expressions can be simplified
in the $T\rightarrow0$ limit analogously to the earlier investigated case of antiferromagnetic order \cite
{Schmalian,QS,Katanin}.

\begin{figure}[tbp]
\noindent\center\includegraphics[width=0.5\textwidth]{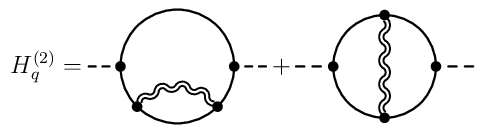}
\caption{\textsf{{\protect\footnotesize {The diagrams for the polarization operator $ H_q$ in the lowest
order in $I$, the notations are the same as in Fig. 2}}}}
\label{chi-st}
\end{figure}

Let us consider for example the lowest (second) order result of perturbation
theory in $I$ for the polarization operator susceptibility $H_q$. There are two topologically different diagrams in this
order, see Fig. \ref{chi-st}.
The corresponding analytical expression reads (cf. Appendix B)
\begin{equation}
H^{(2)}_q=-\frac{6T^2I^2}{N^2}\sum_{kq^{\prime}}\mathcal{G}%
^2_{k}\mathcal{G}_{k+q}\mathcal{G}_{k+q^{\prime}}\mathcal{R}_{q^{\prime}}+\frac{T^2I^2}{N^2}\sum_{kq^{\prime}}\mathcal{G}_{k}\mathcal{G%
}_{k+q}\mathcal{G}_{k+q^{\prime}}\mathcal{G}_{k+q+q^{\prime}}\mathcal{R}_{q^{\prime}},  \label{2ST}
\end{equation}
We assume that the paramagnon propagator $\mathcal{R}_q$ at small momenta and
frequencies for the Fermi level lying near VHS has a form (see, e.g., Ref. \cite{Guinea})
\begin{equation}
\mathcal{R}_{\mathbf{q},\mathrm{i}\omega_n}=\frac{A}{\mathbf{q}%
^2+\xi^{-2}+B|\omega_n|/\max (t q_{+}q_{-},|\omega_n|)}+\mathcal{R}^{\mathrm{r}}_{%
\mathbf{q},\mathrm{i}\omega_n},  \label{Rq}
\end{equation}
where $A, B>0$ are some constants, $\xi$ is the correlation length of
spin fluctuations, $\mathcal{R}^{\mathrm{r}}_{%
\mathbf{q},\mathrm{i}\omega_n}$ is the
contribution, which is regular at $q\rightarrow0$, $\xi\rightarrow\infty$ and neglected below. The result (\ref{Rq}) can be
obtained from the general formula (\ref{chiHq}) under the assumption that the
expansion of the polarization operator $H_q$ in the
momentum and frequency has the same form as one for the
noninteracting susceptibility $\chi_q^0$, the effect of
the interaction is in the change of the parameters $A$ and $B$ only. While the condition $B>0$ being a consequence of the analytical
properties of the susceptibility as a function of frequency, the
positivity of the constant $A$ is analyzed below in the section 3.1.

It follows from the Eq. (\ref{Rq}) that the dominant contribution to the momenta- and frequency sum in Eq. (\ref{2ST})
comes from the region with $|\mathbf{q}|\sim\xi^{-1}$ and the zeroth Matsubara frequency. If the
condition
\begin{equation}
(t/T)^{1/2}\ll\xi  \label{Uu}
\end{equation}
is fulfilled the contribution of the terms with nonzero Matsubara
frequency in the sum over $\omega^{\prime}$ in the Eq. (\ref{Rq}) can be neglected, together with the $%
\mathbf{q}$-dependence of the electronic Green function. The condition (\ref{Uu}) is certainly fulfilled in 2D case at
finite $T$ above the ordered ground state, since of the
correlation length is exponentially large at small $T$ (cf., for example,
Refs. \cite{TPSC,KKI}). This differs present theory from the earlier
considered 3D case \cite{Klenin}, where above-discussed approximations give only
qualitatively, but not quantitatively correct description of magnetic properties. 

Within these approximations the sum (\ref{2ST}) takes a form
\begin{equation}
H^{(2)}_q=\frac{T^2I^2}{N^2}\sum_{k}\left(-6\mathcal{G}%
^3_{k}\mathcal{G}_{k+q}+\mathcal{G}^2_{k}\mathcal{%
G}^2_{k+q}\right) \sum_{\mathbf{q}^{\prime}}\frac1{\omega_{\mathbf{q}%
^{\prime}}}.  \label{2ST2}
\end{equation}
Analogous approximations are also applicable to higher order diagrams.
The results coincide with those for the model
with the action
\begin{equation}
\mathcal{Z}_{\xi\rightarrow\infty}[\eta,\eta^+]=\int d^3S \exp\left[-\frac{%
3I^2S^2}{2\Delta^2}-\beta\sum_{k}{\eta}_{k}^+(\mathrm{i}\nu_n-\varepsilon_{%
\mathbf{k}}-I\vec{\sigma}\mathbf{S})^{-1}\eta_{k}\right],  \label{xi}
\end{equation}
which contains only one uniform static mode $\mathbf{S}$ with the propagator
\begin{equation}
\frac13\left(\Delta/I\right)^2\equiv \frac{T}{N}\sum_\mathbf{q}\frac1{{%
\omega_\mathbf{q}}}.
\end{equation}

The Green functions of Bose and Fermi fields of the model (\ref{xi}) can be expressed
in closed analytical form, cf. \cite{QS,Katanin}.
The static approximation and the neglection of $\mathbf{q}$-dependence of
electronic Green function leads to zero momentum- and frequency transfer
along the paramagnon lines in all diagrams; hence, $\mathbb{G}_{kk^{\prime}}[%
\mathbf{S}]$ becomes diagonal with respect to momenta and frequency:
\begin{equation}
\mathbb{G}_{kk^{\prime}}[\mathbf{S}]\rightarrow G_k(\mathbf{S}%
)\delta_{kk^{\prime}},\; G_k(\mathbf{S})=\left(\mathrm{i}\omega_n-%
\varepsilon_{\mathbf{k}}-I\vec{\sigma}\mathbf{S}\right)^{-1}.\
\label{Replace}
\end{equation}
Therefore we obtain
\begin{equation}
H_q=\left\langle \Pi_q(\mathbf{S}%
)\right\rangle_{\xi\rightarrow\infty},  \label{irr}
\end{equation}
where
\begin{equation}
\Pi_q(\mathbf{S})=-\frac{T}{2N}\sum_{k} \mathrm{Tr}_\sigma \left[G_{k}(%
\mathbf{S})\sigma^zG_{k+q}(\mathbf{S})\sigma^z\right],
\end{equation}
the subscript $\xi\rightarrow\infty$ corresponds to averaging with the
functional (\ref{xi}). Calculation of the average in the Eq. (\ref{irr}) leads to  
\begin{equation}
H_q= \int d^3%
\mathbf{S}\left[\frac13\Pi_{\parallel}(\mathbf{q}|S\, \mathrm{sign}%
\,S^z)+\frac23\Pi_{\perp}(\mathbf{q}|S\, \mathrm{sign}\,S^z)\right]%
\exp\left(-\frac{3I^2S^2}{2\Delta^2}\right),  \label{G}
\end{equation}
where
\[
\Pi_{\parallel,\perp}(\mathbf{q}|S)=-\frac1{N}\sum_{\mathbf{k}}\frac{%
f(\varepsilon_{\mathbf{k}}-IS)-f(\varepsilon_{\mathbf{k}+\mathbf{q}}\mp IS)}{%
\varepsilon_{\mathbf{k}}-\varepsilon_{\mathbf{k}+\mathbf{q}}-IS\pm IS}.
\]
Similarly we find the electronic Green function:
\begin{equation}
G_k=\frac{\delta^2 \mathcal{Z}_{\mathrm{sf}}[\eta,\eta^+]}{\delta \bar{\eta}%
^+_k \delta \eta_k}=\left\langle G_k(\mathbf{S})\right\rangle_{\xi%
\rightarrow\infty}.  \label{G1}
\end{equation}
The explicit expression for $G_k$ was obtained earlier in the paper \cite{Katanin}. The
corresponding spectral function
\begin{equation}
A(\mathbf{k},\omega)=-\frac1{\pi}\mathrm{Im}G_{\mathbf{k}}(\omega)=\frac9{%
\sqrt{6\pi}\Delta^3}(\omega-\epsilon_{\mathbf{k}})^2 \exp\left[-\frac{%
3(\omega-\epsilon_{\mathbf{k}})^2}{2\Delta^2}\right]  \label{A}
\end{equation}
has two-peak (non-quasiparticle) structure at the Fermi surface, which
destroys the quasiparticle picture due to strong ferromagnetic
fluctuations (see Fig. \ref{AA} below). As it was discussed in the paper \cite%
{KKI}, the corresponding violation of Fermi-liquid behavior corresponds to
a quasi-splitting of the Fermi surface at low temperatures, which is related to the
change of the the electronic spectrum in the vicinity of the
magnetically ordered ground state.

\section{The Results for the Phase Diagram, Paramagnon Vertices and
Spectral Functions}

\subsection{The irreducible susceptibility} 

\begin{figure}[tbp]
\noindent \includegraphics[width=0.3\textwidth,clip]{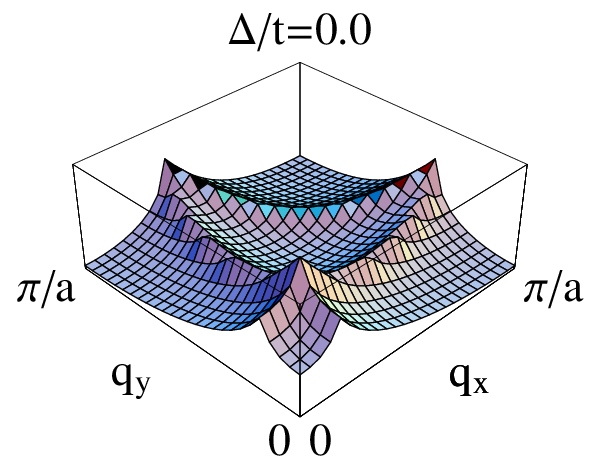} %
\includegraphics[width=0.3\textwidth,clip]{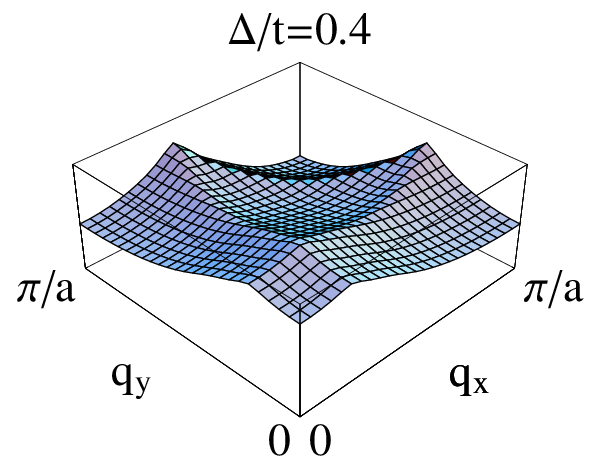} %
\includegraphics[width=0.3\textwidth,clip]{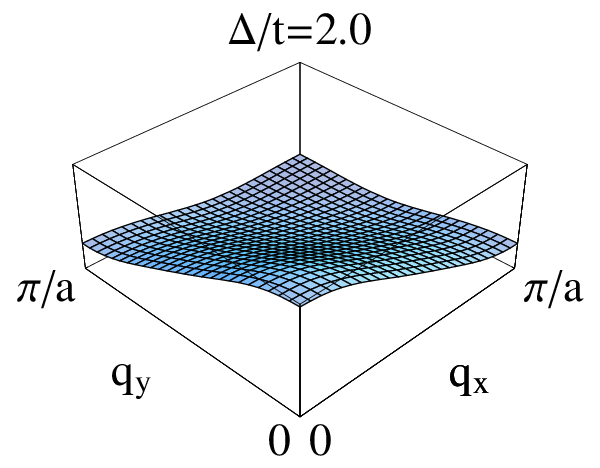}\newline
\vspace{0.5cm}
\caption{{\protect\footnotesize \textsf{Momentum dependence of the static
irreducible electronic susceptibility $\protect H_\mathbf{%
q}$ in the first quarter of the Brillouin zone for $t^{\prime}/t=0.45, n=0.583>n_{%
\mathrm{VH}}=0.466$ and different values of $\Delta$ }}}
\label{q-chi}
\end{figure}

\begin{figure}[tbp]
\noindent \includegraphics[width=0.3\textwidth,clip]{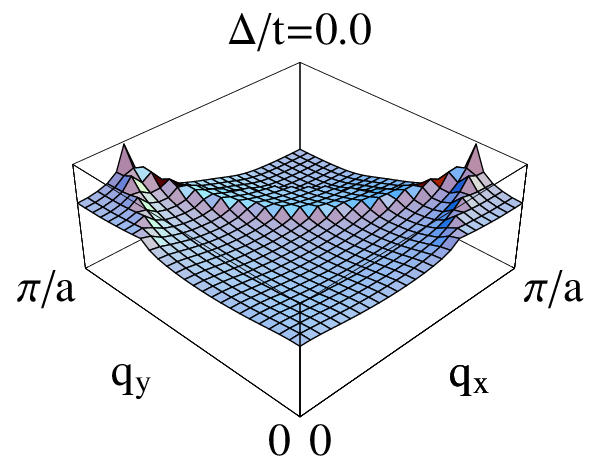} %
\includegraphics[width=0.3\textwidth,clip]{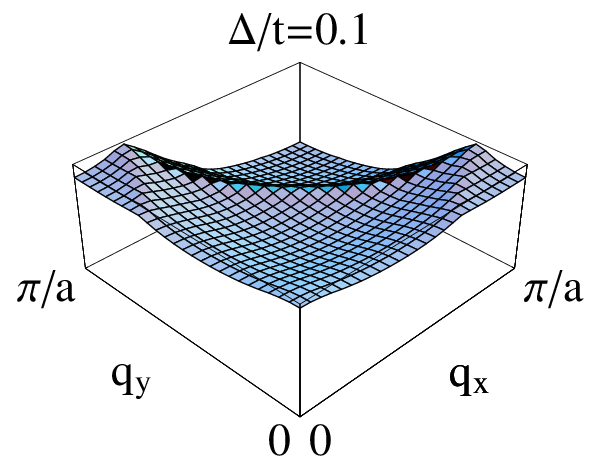} %
\includegraphics[width=0.3\textwidth,clip]{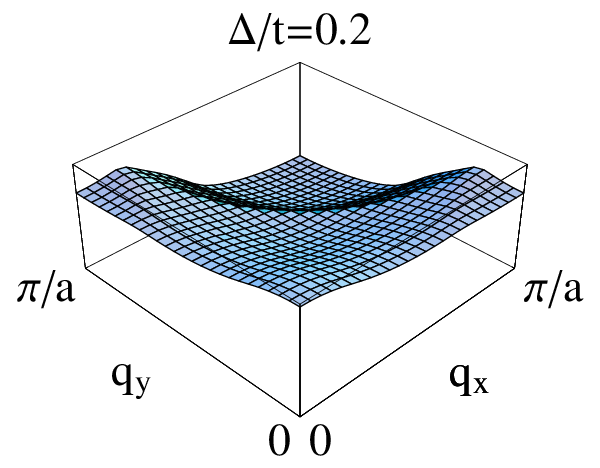}\newline
\vspace{0.5cm}
\caption{{\protect\footnotesize \textsf{The same as in Fig. \protect\ref%
{q-chi}, for $n=0.338<n_{\mathrm{VH}}$ }}}
\label{q-chi2}
\end{figure}

To investigate the possibility of the ferromagnetic order in the ground
state we consider the momentum dependence of the irreducible static susceptibility (polarization
operator) $H_{(\mathbf{q},0)}$ at $T\rightarrow0$. The plots of $H_{(\mathbf{q},0)}$, calculated according to the Eq. (\ref{irr}) in
a quarter of the Brillouin zone for $t^{\prime}=0.45t$ and different values of $%
n, \Delta$ are presented in Figs. \ref{q-chi}, \ref{q-chi2}. The
chemical potential $\mu$ is adjusted to keep the number of electrons $%
n=(2/N)\sum_{\mathbf{k}}\int_{-\infty}^{\mu}d\varepsilon\,A({\mathbf{k}}%
,\varepsilon)$ equal to the noninteracting value with increasing $\Delta$, the spectral function $A({\mathbf{k}}%
,\varepsilon)$ is determined by the Eq. (\ref{A}).

For $\Delta=0$ (which corresponds to $I=0$) the global maximum of the static
susceptibility is located at the point $\mathbf{q}\ne 0$, so that for both
the positions of the Fermi level above and
below the VHS the condition of the
positivity of paramagnon excitation spectrum is violated. 
With increasing $\Delta$ the static irreducible susceptibility has a
maximum at $\mathbf{q}=0$: for $n<n_{\mathrm{VH}}$ the local maximum at $%
\mathbf{q}=0$ becomes the global one and for $n>n_{\mathrm{VH}}$ local
maximum shifts to the point $\mathbf{q}=0$. Therefore, in both cases the
interaction of electrons with paramagnons leads to the global maximum of
polarization operator of the electronic subsystem at $\mathbf{q}=0$,
which corresponds to a possibility of the ferromagnetically ordered
ground state.

\begin{figure}[tbp]
\center 
\includegraphics[width=0.5\textwidth]{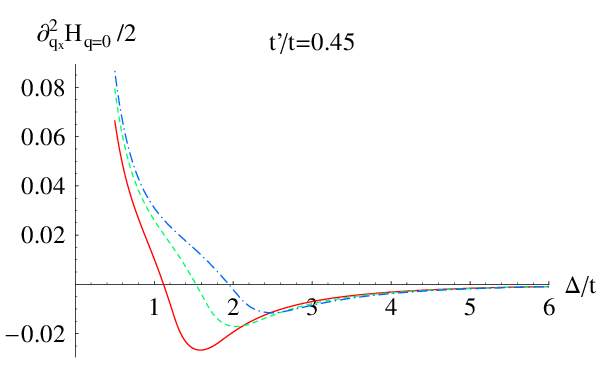} \vspace{0.5cm}
\caption{{\protect\footnotesize \textsf{The derivative $\partial^2_{q_x}%
\protect H_{\mathbf{q}=0}$ vs $\Delta$ for: $n=0.533$ ---
solid line, $n=0.583$ --- dashed line, $n=0.626$ --- dash-dotted line}}}
\label{Der}
\end{figure}

To determine the critical value of $\Delta$ (the minimal value of $\Delta$,
at which the ferromagnetic ground state is possible) we consider the second
derivative of the irreducible susceptibility with respect to $q_x$ (or $q_y$) $%
\partial^2_{q_x}H_{\mathbf{q}}\equiv{\partial^2H_{%
\mathbf{q}}}/{\partial{q^2_x}}$ at the point $\mathbf{q}=0$
(cf. Appendix C). The results of calculation of this are presented in Fig. %
\ref{Der}. Changing of sign of the second derivative determines the critical
values $\Delta_c\sim t$, which depend on the electronic concentration.

\begin{figure}[tbp]
\noindent\includegraphics[width=0.5\textwidth]{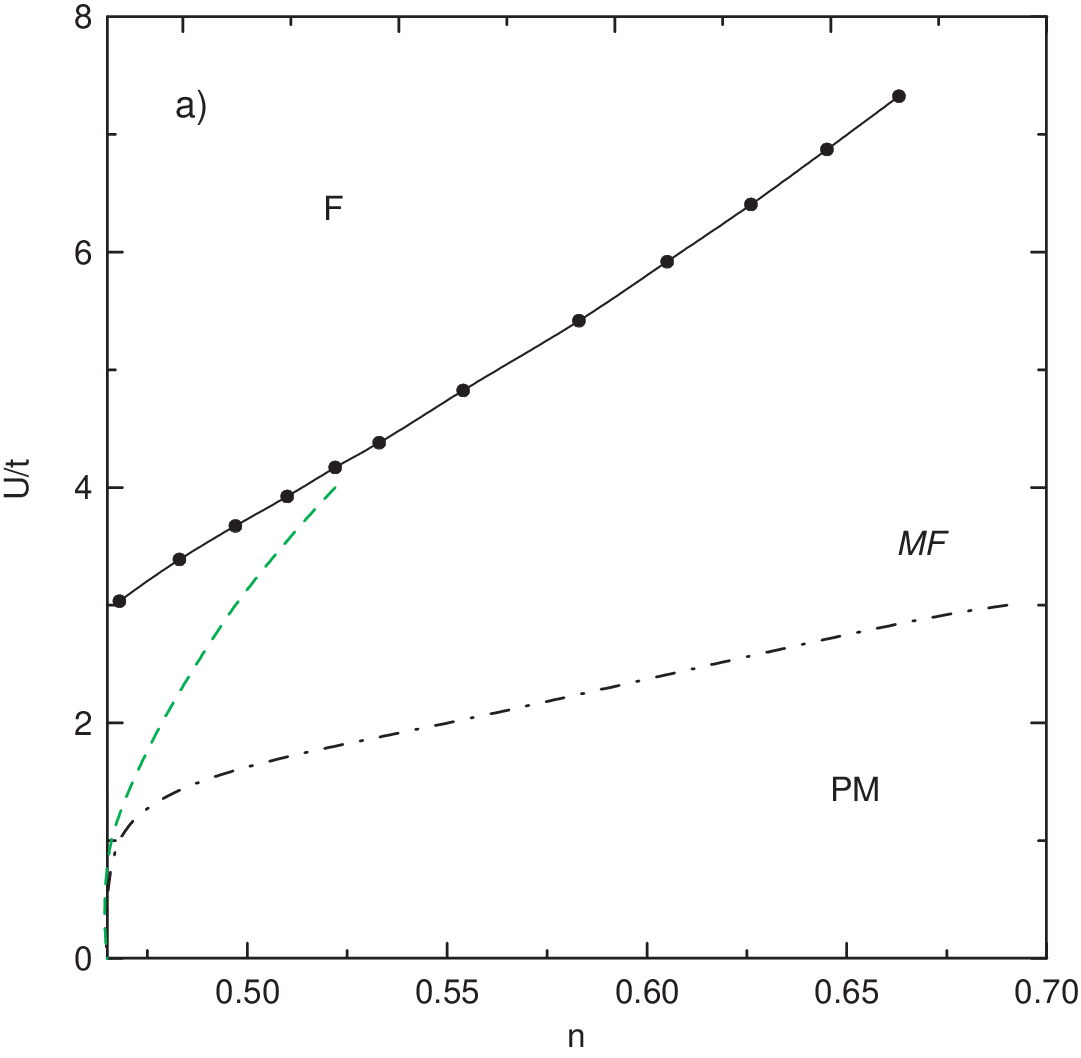} %
\includegraphics[width=0.5\textwidth]{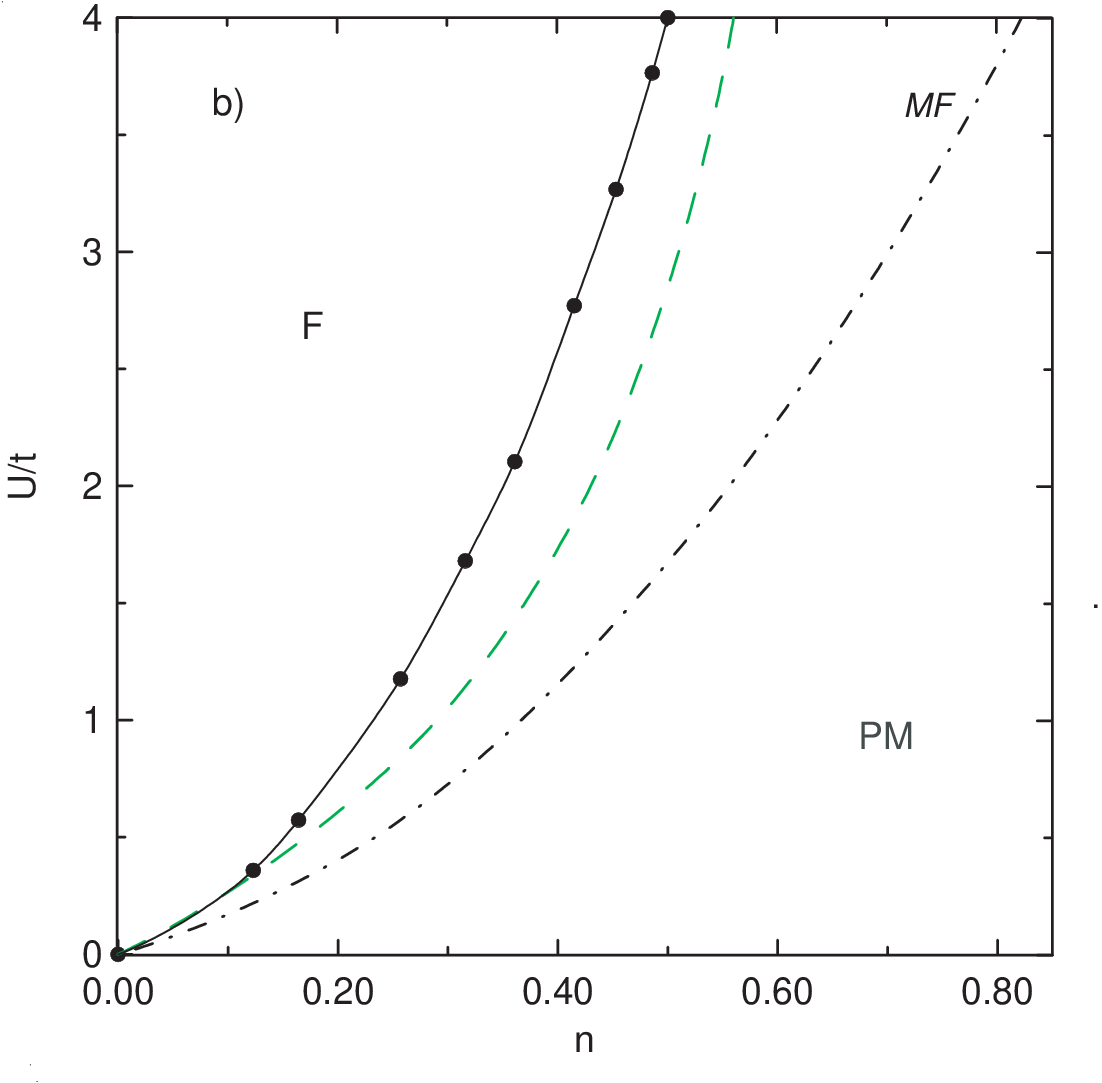} \vspace{0.5cm}
\caption{{\protect\footnotesize \textsf{Phase diagram in $n-U$ plane:
(a) $t^{\prime}/t=0.45$; (b) $t^{\prime}/t=0.50.$ Solid line --- the result
of the present paper; dashed line --- the results of functional renormalization approach
methods \protect\cite{Kampf_RG}; dash-dotted
line --- the result of the mean field theory}}}
\label{phase}
\end{figure}

\subsection{The phase diagram of the Hubbard model}
To obtain the phase diagram it is necessary to establish the
relation between the critical value of parameter $\Delta_c$ and the corresponding
spin-fermion interaction $I_c$ or the Hubbard interaction $U_c$. These relations can be found using the
generalized Stoner criterion (\ref{Stoner}). Below we consider the Hubbard model, for which we find $U_c=1/H_{\mathbf{0}%
}(\Delta_c),$ where $H_{\mathbf{0}}$ is the uniform
static polarizaton operator,
\begin{equation}
H_{\mathbf{0}}=\frac{1}{2\Delta}\int\limits_{-4+8t^{%
\prime}}^{4+8t^{\prime}} \rho(\varepsilon) \varphi\left(\frac{\varepsilon-\mu%
}{\Delta}\right)d\varepsilon,  \label{Hi_irr1}
\end{equation}
$\varphi( x )=(3 x^2 +2) \exp(-{3 x^2}/2)/\sqrt{6\pi}.$

The resulting phase diagram in the $n-U$ plane for $t^{\prime}/t=0.45$ and $n>n_{\mathrm{VH}}$
is presented in Fig. \ref{phase}a (the case $n<n_{\mathrm{VH}}$ is not
considered in the following, because the incommensurate spin fluctuations
with the wavevector $\mathbf{Q}$ far from the point $\mathbf{q}=0$ are not taken
into account in the present approach). The critical values $U_c$ for $n>n_{%
\mathrm{VH}}$ are larger than the corresponding mean-field values $U_c^{%
\mathrm{MF}}$. The results for $n$ far from $n_{\mathrm{VH}}$ are in
qualitative agreement with the functional renormalization group (fRG) results for
the Hubbard model \cite{Kampf_RG}. However, in contrast to the fRG results
the critical value $U_c$ is nonzero for $n=n_{\mathrm{VH}}$, which
is related to the breakdown of the quasiparticle picture of the electronic
spectrum owing to the ferromagnetic fluctuations. Thus, the non-Fermi-liquid
properties of the electronic spectrum in 2D systems become important for
the criterion of the ferromagnetism in the vicinity
of the VHS.

The phase diagram for the flat band case $\;t^{\prime}/t=0.50$ is presented
in Fig. \ref{phase}b. In this case the critical values $U_c$ are also
larger than the mean field values $U_c^{\mathrm{MF}}$, and there is
qualitative agreement between the fRG and obtained here results for all values
of electronic concentration $n$ \cite{Kampf_RG}.


\subsection{The paramagnon interaction vertex}

Let us consider the vertex $\Gamma^{jjj^{\prime}j^{\prime}}(q_1,q_2,q_3,q_4)$,
which determines the energy of the paramagnon interaction and defined by the Eq. (\ref{Gamma_def}) of Appendix B.
Below we are interested in $j\neq j^{\prime}$-the
component of this vertex. This component can be
expressed through the corresponding 2-particle irreducible vertex $\Gamma_{%
\mathrm{irr}}^{jjj^{\prime}j^{\prime}}(q_1,q_2,q_3,q_4)$, which is defined such that it cannot be
split into two connected parts by removing two paramagnons lines. In
general the relation between these two vertices is determined by the parquet set of
diagrams.

\begin{figure}[tbp]
\noindent\center\includegraphics[width=0.8\textwidth]{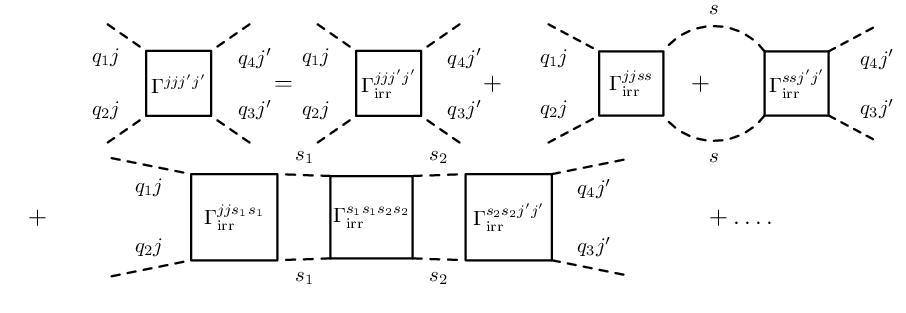}
\caption{\textsf{{\protect\footnotesize {The ladder diagrams for the
paramagnon vertex}}}}
\label{ir-Gamma}
\end{figure}

To simplify the relation between the vertices we generalize the model (\ref{Sc})
introducing the {$M$-component} spin field $\mathbf{S}$ (cf. Refs. \cite%
{Ma,Irk}), $j=1,\ldots,M$. %
Due to internal sums over spin indices, the dominant
contribution into the full vertex in the limit $M\rightarrow\infty$ comes from the ladder set of diagrams
with equal spin indices of internal paramagnon Green functions (see Fig. %
\ref{ir-Gamma}).

This allows to obtain a simple relation between the full and irreducible vertex,
 $\Gamma^{jjj^{\prime}j^{\prime}}$ and $\Gamma_{%
\mathrm{irr}}^{jjj^{\prime}j^{\prime}}$ to first order in $1/M$

\[
\Gamma^{jjj^{\prime}j^{\prime}}(q_1,q_2,q_3,q_4)=\Gamma_{\mathrm{irr}%
}^{jjj^{\prime}j^{\prime}}(q_1,q_2,q_3,q_4)
\]
\begin{equation}
-\frac{T}{8N}\sum_{j^{\prime\prime},p_1,p^{\prime}}\Gamma^{jjj^{\prime%
\prime}j^{\prime\prime}}(q_1,q_2,p,p^{\prime})\mathcal{R}_{p}\mathcal{R}_{p^{\prime}}\Gamma_{\mathrm{irr}}^{j^{\prime\prime}j^{\prime%
\prime}j^{\prime}j^{\prime}}(-p,-p^{\prime},q_3,q_4)\delta_{q_1+q_2,-p-p^{%
\prime}}.  \label{Ladder}
\end{equation}
At large values of the correlation length $\xi$ the dominant contribution
to the sum (\ref{Ladder}) comes from the small momenta $p,p^{\prime}$.
Supposing that the external momenta $q_1,q_2,q_3,q_4$ are also small and $\Gamma_{%
\mathrm{irr}}^{jjj^{\prime}j^{\prime}}(q_1,q_2,q_3,q_4)$ is constant in the vicinity of $\mathbf{q}
_1=\mathbf{q}_2=\mathbf{q}_3=\mathbf{q}_4=0$ (below we denote this constant
as $\Gamma_{\mathrm{irr}})$ we obtain in the framework of the quasistatic
approach
\begin{equation}
\Gamma^{jjj^{\prime}j^{\prime}}(q_1,q_2,q_3,q_4)=\frac{\Gamma_{\mathrm{irr}}%
}{1+\Gamma_{\mathrm{irr}}M\zeta_{\mathbf{q}_1+\mathbf{q}_2}/2}
\delta_{\omega_1,0}\delta_{\omega_2,0}\delta_{\omega_3,0}\delta_{\omega_4,0}
,  \label{Gamma1}
\end{equation}
where
\begin{equation}
\zeta_{\mathbf{q}}=\frac{T}{4N}\sum_{\mathbf{p}}\mathcal{R}_{\mathbf{p}%
}\mathcal{R}_{\mathbf{q-p}}=\frac{T A^2}{4\pi |\mathbf{q}| \sqrt{%
\mathbf{q}^2+4 \xi^{-2}}}\ln \frac{|\mathbf{q}|+\sqrt{\mathbf{q}^2+4 \xi^{-2}%
}}{2 \xi^{-1}}.
\end{equation}
Note, that in the limit $\xi \rightarrow \infty$ the quantity $\zeta_{%
\mathbf{q}}=T A^2/(4\pi |\mathbf{q}|^2)\ln(|\mathbf{q}|\xi)$ diverges at $|%
\mathbf{q}|\rightarrow 0$, which results in vanishing of the full vertex
of paramagnon interaction (\ref{Gamma1}) at $\mathbf{q}_1=\mathbf{q}_2=%
\mathbf{q}_3=\mathbf{q}_4=0$ in accordance with the assumption of existing
of long-range order in the ground state \footnote{%
Actually this vanishing of the vertex generalizes the Adler principle \cite%
{Adler} for the electronic system interacting with paramagnons; this is similar to vanishing of the magnon-interaction vertex at $q_i=0$ in Heisenberg magnets which was
discussed earlier \cite{Adler1}.}.

\begin{figure}[tbp]
\noindent\includegraphics[width=0.5\textwidth,clip]{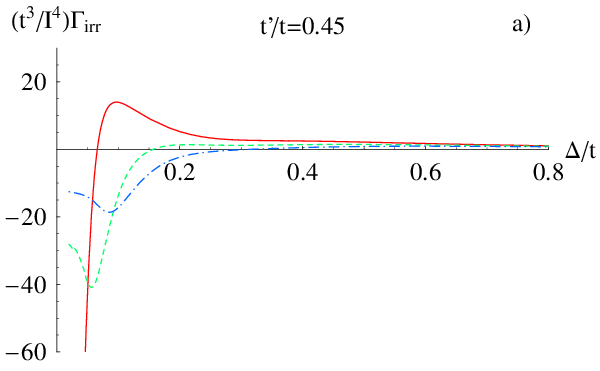}
\noindent\includegraphics[width=0.5\textwidth,clip]{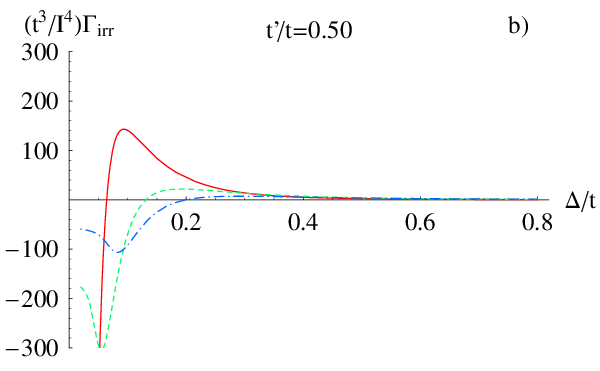}
\caption{\textsf{{\protect\footnotesize The irreducible vertex $%
(t^3/I^4)\Gamma_{\mathrm{irr}}$ vs $\Delta$ at different electronic concentration $n$ and $t'$: (a) $t^{\prime}=0.45t$: $%
n=0.533 $ --- solid line, $n=0.583$ --- dashed line, $n=0.626$ ---
dot-dashed line; (b) $t^{\prime}=0.5t$: $n=0.316$ --- solid line, $n=0.415$
--- dashed line, $n=0.486$ --- dot-dashed line}}}
\end{figure}

One can see from the Eq. (\ref{Gamma1}), that the sign of $\Gamma_{\mathrm{irr}}$ at
small external momenta is crucial for the possibility of the FM ground state. If $\Gamma_{\mathrm{irr}}<0$, the
expression (\ref{Gamma1}) shows that paramagnons can form bound states, and
hence the ferromagnetism is not possible. Let us consider the sign of $%
\Gamma_{\mathrm{irr}}$ within the quasistatic approach depending on the
magnitude of the electron-paramagnon interaction. In general, the vertex $%
\Gamma_{\mathrm{irr}}$ can be represented diagrammatically as a sum of the
contributions with the one-loop vertices $\Gamma_{\mathrm{1-loop}}$, connected by
three or more paramagnon lines. Using the same approximation as for the one-particle polarization operator $\Gamma_{\mathrm{irr}}=\Gamma_{\mathrm{1-loop}},$ i. e. neglecting
contributions to the irreducible vertex with the number of loops $n>1$ we obtain in the $M=3$ case
\begin{equation}
\Gamma^{\mathrm{irr}}=\frac{TI^4}{4N}\mathrm{Tr}_\sigma\left\langle
\sum_kG_k(\mathbf{S})\sigma^zG_k(\mathbf{S})\sigma^xG_k(\mathbf{S}%
)\sigma^zG_k(\mathbf{S})\sigma^x\right\rangle_{\xi\rightarrow\infty}
\label{G_irr}
\end{equation}
\[
\ \ =\frac{1}{\Delta^3}\int_{-4+8t^{\prime}}^{4+8t^{\prime}}
\rho(\varepsilon) g\left(\frac{\varepsilon-\mu}{\Delta}\right)d\varepsilon,
\ \ \ g( x )=\frac{1}{15}(2 + 3 x^2 -9 x^4) \exp\left(-3 x^2/2\right).
\]
The results of the numerical calculation of the irreducible vertex in the $T\rightarrow0$ limit using the Eq. (%
\ref{G_irr}) are presented in Fig. 9. One can see
that the irreducible vertex $\Gamma_{\mathrm{irr}}$ changes its sign at the values of ${\Delta_c}$, which are much smaller than the values of $\Delta_c$, obtained in Section 3.1, and hence they do not result in
additional limitations on the possibility of the ferromagnetic ground state.

For $\Gamma_{\mathrm{irr}}>0$ and small momenta $q_1,\ldots,q_4$ one can neglect the
unity in the denominator of (\ref{Gamma1}) in the limit $%
M\rightarrow\infty$ to obtain
\begin{equation}
\Gamma^{jjj^{\prime}j^{\prime}}(q_1,q_2,q_3,q_4)=\frac2{M\zeta_{\mathbf{q}_1+%
\mathbf{q}_2}}
\delta_{\omega_1,0}\delta_{\omega_2,0}\delta_{\omega_3,0}\delta_{\omega_4,0}.
\label{Gamma2}
\end{equation}
Note, that for $\Gamma_{\mathrm{irr}}\ll 1$ (i.e. in regime of weak coupling
and/or smooth enough density of states) the formula (\ref{Gamma2}) is correct in
the considered limit of low temperature ($\xi\rightarrow\infty$). For higher
temperature the unity in the denominator (\ref{Gamma1}) has to be retained
also in the limit of large $M$, since $\Gamma_{\mathrm{irr}} \zeta_{0}\ll 1$.

\subsection{The electronic self-energy and spectral functions}

The first order $1/M$ result (\ref{Gamma2}) for the paramagnon vertex can be
used to investigate the influence of the paramagnon interaction on the
electronic self-energy $\Sigma$ and other properties of the electronic
system. In the framework of the $1/M$-expansion it was shown \cite{Irk} that the
electronic properties of the model (\ref{Sc}) to first order in $1/M$ are
determined by the self-energy $\Sigma_k$ and
one-particle irreducible vertices of electron-paramagnon interaction $%
\gamma_k\,\gamma_k^{zz},\gamma_k^{zz \perp}$. In the limit $%
\xi\rightarrow\infty$ these quantities can be obtained from the system of the algebraic equations

\[
\Sigma_k=\Delta^2\gamma_kG_k,
\]
\[
\gamma_k=1+(\gamma_k^{zz}G_k-\gamma_k^2G_k^2)\Delta^2/3,
\]
\begin{equation}
\gamma_k^{zz}=\Delta^2(2\gamma^{3}_kG_k^3+\gamma_k\gamma_k^{zz}G_k^2+%
\gamma_k^{zz \perp}G_k)+\alpha\gamma_kG_k,  \label{eqs}
\end{equation}
\[
\gamma_k^{zz
\perp}=-2\Delta^2(\gamma_k^4G_k^4+\gamma_k^2\gamma_k^{zz}G_k^3+\gamma_k%
\gamma_k^{zz \perp}G_k^2)-\alpha\gamma_k^2G_k^2.
\]
where $\alpha=M\Gamma^{jjj^{\prime}j^{\prime}}(0,0,0,0)\zeta_{\mathbf{0}}$.
The results of the solution of these equations with account of paramagnon
interaction ($\alpha=2$ from (\ref{Gamma2})) and without it ($\alpha=0$) are
presented in Figs. 6-8.

\begin{figure}[tbp]
\begin{center}
\includegraphics[width=0.5\textwidth,clip]{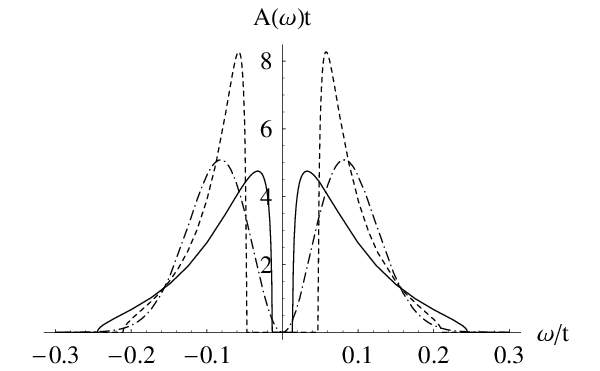}
\end{center}
\caption{{\protect\footnotesize \textsf{The spectral density $A=-\frac1{%
\protect\pi}\Im G$ vs $\protect\omega\equiv\protect\nu-\protect%
\varepsilon_{\mathbf{k}}+\protect\mu$, calculated from (\protect\ref{eqs}) without ($\protect%
\alpha=0$, dashed line) and with account of ($\protect\alpha=2$, solid line)
the paramagnon interaction ; dash-dotted line --- the same in the
quasistatic approach (the formula (\protect\ref{A})), $\Delta=0.1t$}}}
\label{AA}
\end{figure}
\begin{figure}[tbp]
\noindent\includegraphics[width=0.5\textwidth,clip]{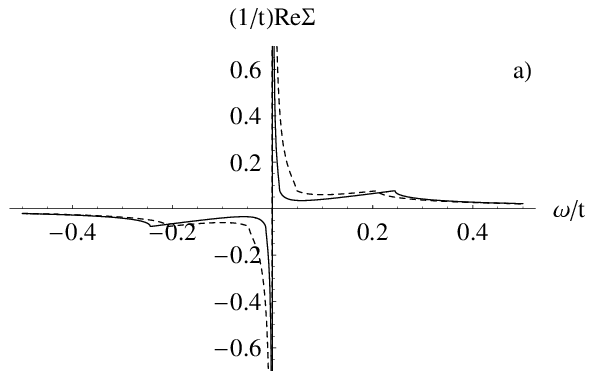} \noindent%
\includegraphics[width=0.5\textwidth,clip]{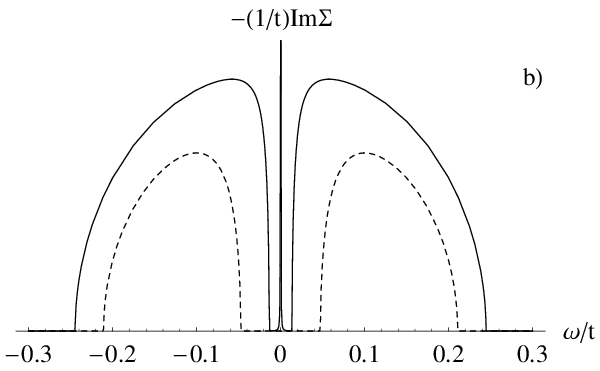}
\caption{{\protect\footnotesize \textsf{The real (a) and imaginary (b) part
of the electronic self-energy $\Sigma(\protect\omega),$ the notations and
the parameters are the same as in Fig. \protect\ref{AA}}}}
\label{SS}
\end{figure}
\begin{figure}[tbp]
\center\noindent\includegraphics[width=0.5\textwidth,clip]{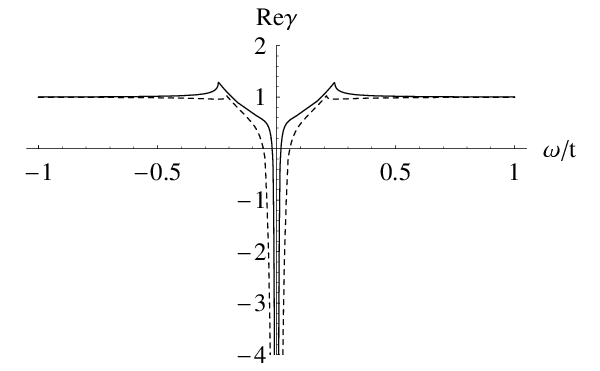}
\caption{{\protect\footnotesize \textsf{The real part of the electron
paramagnon-vertex $\protect\gamma(\protect\omega)$, the notations and the
parameters are the same as in Figs. \protect\ref{AA}, \protect\ref{SS} }}
}
\label{g}
\end{figure}
As well as in the absence of the paramagnon interaction \cite{Katanin}, the
real part of the self-energy has an infinite slope at the Fermi level, and
imaginary part has a $\delta $-function peak. It was pointed out above that
these anomalies originate from the electron-paramagnon interaction and
violate the quasiparticle picture in the vicinity of the Fermi level.
Physically this corresponds to appearance of the new quasi-split-Fermi-surfaces
shifted from the Fermi level by $\pm \Delta$ in the presence of strong
ferromagnetic fluctuations \cite{KKI}. Note, that existence of the window
where the spectral weight is zero is an artefact of $1/M$-expansion: the
comparison with the results of the quasistatic approach shows (Fig. \ref{AA}%
) that the spectral weight in this window is reduced, but is not zero. In the presence of the paramagnon interaction the
spectrum is less coherent than
without it ($\Im\Sigma$
is enhanced, $A(\omega)$ is smeared), the distance between the peaks of the
spectral density $A$ is being reduced. Thus, account of the interaction is important
for the description of the electronic spectrum.

\section{Conclusions}

We have considered the necessary conditions of the existence of
ferromagnetism in 2D systems in the presence of VHS: the maximum of the
polarization operator at $\mathbf{q}=0$, which ensures positive
definiteness of the magnetic excitation spectrum and the repulsion between the paramagnons (the latter guarantees the impossibility of the bound state formation).


For the Fermi level
position in the vicinity of VHS and in the absence of the electron-paramagnon interaction $I$ these conditions are violated: the momentum
dependence of spin susceptibility has its maximum at the wave vector $%
\mathbf{q}\ne0$, which corresponds to possibility of the spin density
wave in the ground state and the paramagnon interaction is attractive. However, with increasing $I$ the static spin susceptibility acquires a maximum at $\mathbf{q}%
=0\!$ and the paramagnon interaction changes its sign. The critical value of the
electron-paramagnon interaction $I_c$ (or the corresponding Hubbard interaction $U_c$) for which the ferromagnetic ground
state is possible exceeds substantially the corresponding value $I_c^{%
\mathrm{MF}}$ ($U_c^{%
\mathrm{MF}}$), determined from the Stoner theory, and agrees
with the functional renormalization group investigations of Hubbard model
for the electronic concentration being in the vicinity of
the Van Hove filling. The critical values $I_c$ and $U_c$ are however finite at
the Van Hove electronic filling for ${t^{\prime}<0.5t}$, which is related
to the non-quasiparticle picture of the electronic spectrum
(quasi-splitting of the Fermi surface).

In contrast to the Moriya theory, where the difference of the conditions of ferromagnetism from the Stoner
criterion originates from the quantum contributions to the uniform irreducible
susceptibility, in the present approach it results from the change of the
momentum dependence of the susceptibility by classical spin fluctuations.
The change of the uniform irreducible susceptibility by the interaction
is also taken into account in the present
approach, but it is less important. The kind of the transition
from the ferromagnetic to the paramagnetic phase is an open question. One of
the possibilities is the existence of intermediate state with the strong
short-range order, characterized by the wave vector $\mathbf{Q}\ne0$. To
describe magnetic properties of this phase, it is necessary to
consider the excitations with incommensurate wave
vector $\mathbf{Q}$ and account for quantum fluctuations,
which is a problem for further investigations. The problem of
ferromagnetism formation at $n<n_{\mathrm{VH}}$ requires also further consideration.

Besides considered reasons of deviations from the Stoner criterion, the
effect of screening due to particle-particle scattering \cite{Kanamori} may be important in  the Hubbard model. The agreement of the obtained results with the
renormalization group results for this model at fillings away from the
the Van Hove filling shows, however, that the change of momentum dependence of
susceptibility seemingly plays a dominant role. On the other hand, the importance
of the non-quasiparticle spectrum of excitations in the vicinity of
ferromagnetic instability, which has not been taken into account by previous fRG considerations, is demonstrated in the framework of our approach.



The obtained results may serve as a basis for further investigations of
itinerant systems in the vicinity of the ferromagnetic instability in the presence
of VHS in the electronic spectrum. More detailed investigations of the
vicinity of the quantum phase transition and the role of nonanalytic
corrections in the system with presence of VHS are needed. It is also of
interest to investigate a possibility of the triplet pairing in systems with
 of VHS in the electronic spectrum.

\section*{Acknowledgements}

We are grateful to M. I. Katsnelson for valuable discussions. The work was
partially supported by the Project of Scientific Schools of the Russian
Basic Research Foundation 4640.2006.2, 07-02-01264-a and 05-02-217704.


\section*{Appendix A. The Derivation of the Spin-Fermion Model}

In this Appendix we consider the derivation of the spin-fermion model
starting from the Hubbard Hamiltonian (\ref{Hubb}).

The originally
proposed way of derivation of this model \cite{Hertz,Klenin} was to use the
decomposition of the interaction term into charge and spin degrees
of freedom, e.g.
\begin{equation}
Un_{i\uparrow }n_{i\downarrow }=\frac{U}{4}\left[ \left(c_{i%
\sigma }^{+}c_{i\sigma }\right)^{2}\delta _{\sigma \sigma ^{\prime
}}-\left(c_{i\sigma }^{+}\sigma _{\sigma \sigma
^{\prime }}^{x,y,z}c_{i\sigma ^{\prime }}\right)^{2}\right] ,
\label{Decoupl2}
\end{equation}%
(the summation over repeated spin indices is assumed) with the
subsequent decoupling of quartic terms of fermionic operators using auxiliary
scalar or vector fields. The representation (\ref{Decoupl2}) is, however,
not unique, in particular another, $SU(2)$ symmetric representation,
\begin{equation}
Un_{i\uparrow }n_{i\downarrow }=\frac{U}{4}\left[ (c_{i\sigma
}^{+}c_{i\sigma })^{2}\delta _{\sigma \sigma ^{\prime }}-\frac{1}{3}%
(c_{i\sigma }^{+}\sigma _{\sigma \sigma ^{\prime }}c_{i\sigma ^{\prime
}})^{2}\right],  \label{Decoupl1}
\end{equation}%
may be used (see, e.g. Ref \cite{Moriya}).

It was previously discussed (see,
e.g. Refs. \cite{Schulz,Weng, LargeN,Bickers,Dupuis,Wetterich}), that although the
representations (\ref{Decoupl2}) and (\ref{Decoupl1}) are equivalent, they
lead to different effective actions (the so-called Fritz  ambiguity). In particular, the representation (\ref%
{Decoupl2}) leads to an action with the scalar auxiliary fields and therefore is not transparently $SU(2)$ invariant.  The representation (%
\ref{Decoupl1}) is free from this problem, but does not reproduce
correctly the mean-field results for the Hubbard model due to an extra factor of $1/3$.
The reason for this difficulty
is that the factor $1/3$ in Eq. (\ref{Decoupl1}) takes into account that not only a  lowest order component of the longitudinal interaction
$(c_{i\sigma }^{+}\sigma^z _{\sigma \sigma ^{\prime }}c_{i\sigma ^{\prime
}})^2$,
but also the higher order transverse fluctuations which arise from the $(c_{i\sigma }^{+}\sigma^{x,y} _{\sigma \sigma ^{\prime }}c_{i\sigma ^{\prime
}})^2$ part of the interaction contribute to the self-energy of the $S^z$ field and similar to the other quantities (see, e.g., Ref. \cite{Bickers}). 

Below we construct an effective
action which is explicitly $SU(2)$ invariant  and at the same time reproduces correctly the mean-field
results. Although some attempts of derivation of such an action were
performed earlier, they were either restricted to systems with additional
orbital degeneracy \cite{LargeN} or used additional auxiliary vector field to
perform averaging over directions in Eq. (\ref{Decoupl2}) \cite%
{Schulz,Dupuis}.

We start from the Hubbard model action

\begin{equation}
\mathcal{S}_{\mathrm{H}}=\sum\limits_{k\sigma }(-\mathrm{i}\nu
_{n}+\varepsilon_{\bf k})c_{k\sigma }^{+}c_{k\sigma %
}+\mathcal{S}_{\mathrm{int}}-\sum_{k\sigma }\left( \eta _{k\sigma }^{+}c_{k\sigma
}+c_{k\sigma }^{+}\eta _{k\sigma }\right), \label{H1} \end{equation}
\begin{equation}
\mathcal{S}_{\mathrm{int}}=\frac{U}{4N}\sum\limits_{q}(n_{q}n_{-q}-s^z_{q}
s^z_{-q}),\mathbf{s}_{q}=\sum\limits_{k\sigma\sigma^{\prime}}c_{k\sigma }^{+}%
\overrightarrow{\sigma}_{\sigma\sigma^{\prime}}c_{k+q\sigma ^{\prime }},\label{H2}
\end{equation}
where we have introduced fermionic source fields $\eta,\eta^+$.
To derive an effective action we introduce auxiliary field $\mathbf{S}_q$ such that
\begin{equation}
\mathcal{S}_{\mathrm{int}}=\frac1{N}\sum_{q}\left\{ R^{-1}_{q}\mathbf{S}_{q}\mathbf{S}%
_{-q}+U_{\rm eff}\mathbf{s}_{q}\mathbf{S}_{-q}+\frac{U}{4}\left[
n_{q}n_{-q}+U_{\rm eff}\chi
_{q}^{0}\left(
\mathbf{s}_{q}\mathbf{s}_{-q}\right)-(s_{q}^{z}s_{-q}^{z})\vphantom{\frac{U\chi
_{q}^{0}}{1+U\chi
_{q}^{0}}}  \right] \right\} \label{Seff1}
\end{equation}%
where $R^{-1}_{q}=(\chi
_{q}^{0})^{-1}-U_{\rm eff}, U_{\rm eff}=U/(1+U\chi
_{q}^{0}), \chi
_{q}^{0}$ is the bare polarization operator (see Eq. (\ref{Pauli})).
It can be shown that the integration over the field $\mathbf{S}_q$ leads to the original action (\ref{H2}). Let us now consider the sum of the ladder diagrams, generated by the interaction $U_{\rm eff}s^z_{q}S^z_{-q}$. This summation results in the renormalizaton of the $z$-component of the interaction itself $U_{\rm eff}s^z_{q}s^z_{-q}\rightarrow Us^z_{q}s^z_{-q}$, the  electron-paramagnon vertex, $U_{\rm eff}s^z_{q}S^z_{-q}\rightarrow Us^z_{q}S^z_{-q}$ and the renormalization of the paramagnon propagator $R_{q}\rightarrow \chi_q$, where
\begin{equation}
\chi _{q}^{-1}=(\chi
_{q}^{0})^{-1}-U(1-U\chi
_{q}^{0}).\label{chiS}
\end{equation} 
Introducing the two-particle irreducible interaction $(s^z_qs^z_{-q})_{z\rm{-irr}}$, where the index $z$-${\rm irr}$ means that the ladder diagrams generated by this interaction have to be excluded, the result (\ref{Seff1}) can be therefore rewritten in the form

\begin{eqnarray}
\mathcal{S}_{\mathrm{int}} &=&\frac1{N}\sum_{q}\left\{ \chi
_{q}^{-1}S_{q}^{z}S_{-q}^{z}+R^{-1}_{q}(S_{q}^{x}S_{-q}^{x}+S_{q}^{y}S_{-q}^{y})+%
U_{\rm eff}\left( s_{q}^{x}S_{-q}^{x}+s_{q}^{y}S_{-q}^{y}\right)
\right. . \nonumber\\
&&\left. \left. +Us_{q}^{z}S_{-q}^{z}+\frac{U}{4}\left[ n_{q}n_{-q}-(s_{q}^{z}s_{-q}^{z})_{z%
\mathrm{-irr}}\right] +U_{\rm eff}\chi
_{q}^{0}\left(
s_{q}^{x}s_{-q}^{x}+s_{q}^{y}s_{-q}^{y}\right) \right]\right\},\label{HSint}
\end{eqnarray}%
Note, that in addition to the spin-fermion interation the action (\ref{HSint}) contains the Hubbard interaction irreducible in the $s^z$-spin
channel. Performing the same steps in $x$ and $y$ spin directions we obtain
\begin{equation}
\mathcal{S}_{\mathrm{int}}=\frac1{N}\sum_{q}\left\{ \chi _{q}^{-1}%
\mathbf{S}_{q}\mathbf{S}_{-q}+U\mathbf{S}_{q}\mathbf{s}_{-q}\right. 
+\left. \frac{U}{4}\left[ n_{q}n_{-q}-(\mathbf{s}_{q}\mathbf{s}_{-q})_{%
\mathrm{ph-irr}}\right] \right\} \label{App_end}
\end{equation}%
%
where the subscript ph-irr denotes that the ladder diagrams in longitudinal or transversal particle-hole channel  generated  by the interaction $(\mathbf{s}_q\mathbf{s}_{-q})$ are excluded.
Consideration of higher order in $U$ contributions to the ladder-type processes leads to the same action (\ref{App_end}) with the exact polarization operator $H_q$ instead of the bare one $\chi
_{q}^{0}$ in the Eq. (\ref{chiS}).

The effective model (\ref{App_end}) contains both the spin-fermion (first two terms) and the irreducible Hubbard (third term) interaction. Due to its irreducibility the latter is expected to lead only to small renormalization of the parameters of the spin-fermion model and it is neglected in the main text.

To obtain the magnetic susceptibility of the Hubbard model (\ref{H1}) we differentiate the corresponding generating functional (\ref{H1}), (\ref{App_end}) with respect to fermionic sources. In this way we obtain two contibutions:
\begin{equation}
\chi ^{\mathrm{el}}_{q}=\chi ^{\mathrm{el}}_{q,{\,\rm SF}}+U H_q^2,  \label{hiel}
\end{equation}
where the first term originates from the one-particle reducible diagrams containing the spin-fermion interaction $U\mathbf{S}_{q}\mathbf{s}_{-q}$,
\begin{equation}
\chi ^{\mathrm{el}}_{q,{\,\rm SF}}=\frac{H_q}{1-U^2H_q\chi_q},  \label{hiel2}
\end{equation}
and the second term originates from the irreducible Hubbard interaction. The result (\ref{hiel2}) is used in main text, Eq. (\ref{hiel3}).



\section*{Appendix B. The Diagram Technique for the Paramagnon Vertex in the
Spin-Fermion Model}

To consider spin fluctuations in the spin-fermion model (\ref{SR}) it is convenient to
introduce the \textit{$r$-vertex of paramagnon interaction},
\begin{equation}
\Gamma^{j_1\ldots j_r}(q_1,\ldots,q_r)=-2^rT \left (\prod_{i=1}^r
\chi_{q_i}^{-1} \right ) \int\limits_0^\beta d\tau _1\ldots d\tau_r
e^{i\omega_1 \tau _1+...+i\omega_r\tau _r} \\
\langle T_\tau[S_{\mathbf{q}_1}^{j_1}(\tau _1)...S_{\mathbf{q}_r}^{j_r}(\tau
_r)]\rangle_c,  \label{Gamma_def}
\end{equation}
\[
q_1+\ldots+q_r=0,
\]
where $q_1, \ldots, q_r$ are the paramagnon momenta, $j_1, \ldots, j_r$ are the
spin indices. The product $\prod_{i=1}^r \chi_{q_i}^{-1} $ removes the external
Green functions of the field $\mathbf{S}$; for $r>2$ we assume $%
\chi_{q_i} \rightarrow \mathcal{R}_{q_i}$ which corresponds to removing
the exact propagators of the field $\mathbf{S}$, the average with the index $c$
denotes that the diagrams, for which the external spin operators (\ref%
{Gamma_def}) are connected with each other through the lines corresponding
to the electronic Green functions (see below), should be considered. Whereas the 2-vertex
describes self-energy corrections to the paramagnon propagator, the higher order
vertices describe interaction of the renormalized (physical) paramagnons.

The calculation of correlation functions of spin operators can be
carried out using the expansion $\exp[-\beta\mathcal{S}_{\mathrm{int}}]$ of
the functional (\ref{SR}) in the series in electron-paramagnon
interaction. According to the Wick theorem, the average in (\ref{Gamma_def})
is expressed through all possible combinations of averages of pairs of operators
\[
\left\langle S^j_qS^{j^{\prime}}_{q^{\prime}}\right\rangle\equiv
T\int_0^{\beta}d\tau d\tau^{\prime}{ e^{\I\omega_n\tau+\mathrm{i}
\omega_{n^{\prime}}\tau^{\prime}}}\left\langle T_{\tau} \left[S^j_{\mathbf{q}%
}(\tau)S^{j^{\prime}}_{\mathbf{q}^{\prime}}(\tau^{\prime})\right]%
\right\rangle= \frac12N\mathcal{R}_q\delta_{q,-q^{\prime}}\delta_{jj^{\prime}},
\]
where $\mathcal{R}_q$ is the paramagnon propagator. The corresponding
contributions to the vertex can be represented by diagrams,  every
diagram for the $r$-vertex (\ref{Gamma_def}) of the order $n$ consisting of:

\begin{itemize}
\item $n$ internal ($\sigma^{\pm},\sigma^{z}$), $r$ external ($S^{\pm},S^z$)
interaction vertices, and $n$ solid lines, corresponding to
the electronic Green functions. One electronic line enters and goes out of each internal vertex;

\item the longitudinal vertices ($\sigma^{z},S^{z}$) are connected by dashed
lines, the transverse vertices ($\sigma^{\pm},S^{\pm}$) are
connected by wavy lines. The external $S^+$-vertex is
connected with the internal $\sigma^+$-vertex, the $S^-$-vertex with $\sigma^-$%
, and $S^z$-vertex with $\sigma^z$; internal wavy lines connect $%
\sigma^+$- and $\sigma^-$-vertices, and dashed lines $\sigma^z$-vertices;

\item The factor $(1/2)\mathcal{R}_q$ corresponds to each internal dashed or wavy
line with the 4-momentum $q$, the factor $\mathcal{G}_k$ --- to
each solid line with the 4-momentum $k$; the sum of momenta, which
enter each internal vertex, is equal to the sum of momenta going out
of this vertex, the summation over all independent momenta in the
resulting expression being performed;

\item The number of internal transverse vertices is even for each loop and $%
\sigma^+$- and $\sigma^-$-vertices alternate along the electronic Green function line. The type of all transverse vertices in diagram for
which it is possible to determine it unambiguously should be determined;

\item The factor corresponding to a diagram is $I^{n}\left(T/N%
\right)^{(n-r)/2+1}(-1)^{c+f+1}2^{n_\perp+n_K}$,  where $c$ is the full
number of longitudinal vertices, placed between neighboring transverse $%
\sigma^-$- and $\sigma^+$-vertices. $f$ is the number of electronic loops, $n_K$
is the number of the independent internal transverse vertices, type of
which is not determined unambiguously, $n_\perp$ is the number of wavy
lines;
\end{itemize}

The lowest-order diagram for the vertex of any order is
the electronic loop (\ref{Loop}). 
The vertex renormalization is performed in two ways: the
internal paramagnon lines can appear in electronic loop or consist of several loops, connected by paramagnon lines.

To classify diagrams it is convenient to define the \textit{one-loop} vertex
$\Gamma_{\mathrm{1-loop}}^{j_1\ldots j_r}(q_1,\ldots,q_r)$ as a sum of
diagrams for the vertex containing only one electronic loop of electronic Green function
(see Fig. 2b). Every diagram for the vertex consists of one or more one-loop
vertices, connected by paramagnon lines. 
Comparing coefficients of expansions in the series in $I$ of (\ref{Gamma_def}%
) and (\ref{one-loop}), it is possible to show that
\[
\Gamma_{\mathrm{1-loop}}^{j_1\ldots j_r}(q_1,\ldots,q_r)=\frac{T I^{r}}{rN}%
\sum_{\mathcal{P}_q}\mathrm{Tr}_\sigma\sum_{k_1\ldots k_r}
\]
\begin{equation}
\left\langle \mathbb{G}_{k_1,k_2}[\mathbf{S}]{\sigma}^{j_1}\mathbb{G}%
_{k_2+q_1,k_3+q_1}[\mathbf{S}]{\sigma}^{j_2}\ldots\mathbb{G}%
_{k_r+q_1+\ldots+q_{r-1},k_1+q_1+\ldots+q_{r-1}}[\mathbf{S}]{\sigma}%
^{j_r}\right\rangle_{S},  \label{one-loop}
\end{equation}
index $\mathcal{P}_q$ points that the sum is taken over all permutations
numbers $i$ of momenta $q_i$ and \textit{simultaneously} spin indices $j_i$.
The explicit expressions for one-loop 2- and 4-vertices are given by
formulas (\ref{G}) and (\ref{G_irr}) of main text.


\section*{Appendix C. The Second Derivative of the Magnetic
Susceptibility}

The second derivative of the momentum dependent susceptibility at $T=0$ has a
form
\[
\partial^2_{q_x}H_{\mathbf{q}=0}(\omega=0)=\frac{\sqrt{3/8\pi }}{\Delta^2}\times
\]

\begin{equation}
\times
\int\limits_{(w_1-\mu)/\Delta}^{(w_2-\mu)/\Delta}d\xi\,\rho_p(\mu+\xi\Delta)
\left[\left(-3\xi^4+5\xi^2-8/3\right)\exp\left(-3\xi^2/2\right)-\mathrm{Ei}%
\left(-3\xi^2/2\right)\right],  \label{D2}
\end{equation}
where $w_{1,2}=\pm 4t+8 t^{\prime}$ 
, $%
\mathrm{Ei}(x)=\int_{-x}^{+\infty}dt\,e^{-t}/t$, $\rho_p(\varepsilon)$ is
the average value of square of electronic velocity in the iso-energetic
surface $\varepsilon_{\mathbf{k}}+\mu=\varepsilon$,
\begin{equation}
\rho_p(\varepsilon)=\frac1{N}\sum_{\mathbf{k}}\delta(\varepsilon+\mu-%
\varepsilon_{\mathbf{k}})v_{\mathbf{k}}^2=\oint\limits_{\varepsilon_{\mathbf{%
k}}=\varepsilon+\mu} \frac{d\sigma}{(2\pi)^2}v_\mathbf{k}  \label{vks}
\end{equation}
The calculation of the integral in (\ref{vks}) results in
\[
\rho_p(\varepsilon)=\frac{1}{\pi^2(t^2+\varepsilon t^{\prime}-4{t^{\prime}}%
^2)^{1/2}}\times
\]
\[
\times\left[(-2|\varepsilon|t+4t^{\prime}\varepsilon-\varepsilon^2)\mathrm{F}%
(k^2)+8\left(t^2+\varepsilon t^{\prime}-4{t^{\prime}}^2\right)\mathrm{E}(k^2)+%
\frac{\varepsilon^2 t}{t+2t^{\prime}\mathrm{sign}\varepsilon}\Pi\left(1-\frac{%
|\varepsilon|}{4t+8t^{\prime}\mathrm{sign}\varepsilon},k^2\right)\right],
\]
where
\[
k^2=-\frac{(-4t+\varepsilon-8t^{\prime})(4t+\varepsilon-8t^{%
\prime})} {16(t^2+\varepsilon t^{\prime}-4t^{\prime})^2},
\]
and $\mathrm{F}, \mathrm{E}, \Pi$ are the elliptic integrals first, second and
third kind respectively. According to (\ref{D2}) $\partial^2_{q_x}%
H_{\mathbf{q}=0}(\omega=0;\Delta\rightarrow0)=-1/12
\rho_p^{\prime\prime}(\mu).$

\begin{figure}[h]
\noindent\includegraphics[width=0.5\textwidth]{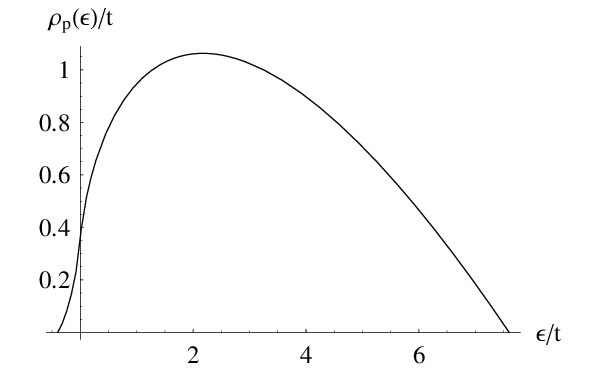} %
\includegraphics[width=0.5\textwidth]{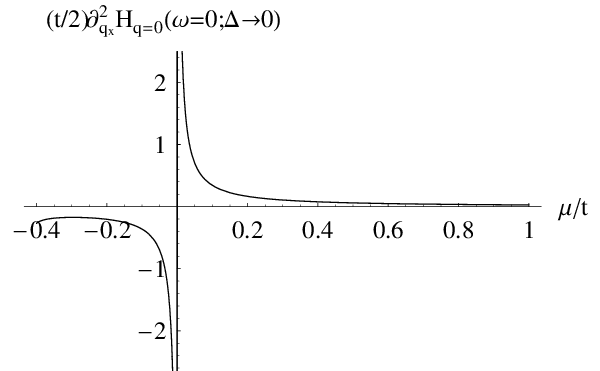} \vspace{0.5cm}
\caption{{\protect\footnotesize \textsf{The prot of : (a) $\protect\rho_p(%
\protect\varepsilon)$;(b) $\partial^2_{q_x}\protect H_{\mathbf{q}=0}(\protect\omega=0;\Delta\rightarrow0)$ vs $\protect\mu $}}}
\label{ro_p}
\end{figure}

Plots of the functions of $\rho_p(\varepsilon)$ and the second derivative of
susceptibility with respect to momentum $\partial^2_{q_x}H_{\mathbf{q}%
=0}(\omega=0;\Delta\rightarrow0)$ are presented in Fig. %
\ref{ro_p}.
The sign of $\partial^2_{q_x}H_{\mathbf{q}=0}(\omega=0;\Delta\rightarrow0)$ coincides with the sign of $\mu$. Hence, the
susceptibility of noninteracting electrons has its local maximum in the
point $\mathbf{q}=0$ for electronic concentration $n<n_{\mathrm{VH}}$ and
minimum for $n>n_{\mathrm{VH}}$. The second derivative of the
susceptibility with respect to momentum tends to plus (minus) infinity as the
electronic concentration approaches the Van Hove filling from the above (below).

\end{document}